\documentclass[onecolumn,secnumarabic,floatfix,amssymb,amsmath,nofootinbib,tightenlines,showpacs,superscriptaddress,aps,prb]{revtex4}
\usepackage{amsfonts}
\usepackage[colorlinks=true,linkcolor=blue,pagecolor=blue,filecolor=blue,menucolor=yellow,urlcolor=blue,citecolor=blue,anchorcolor=blue]{hyperref}%
\usepackage{graphicx}
\usepackage{dcolumn}
\usepackage{times}

\begin{document}
\title{Spin Diffusion Coefficient of $A$ Phase of Liquid $^{3}He$ at
Low Temperatures and Stability of Half Quantum Vortex}

\author{Shokouh Haghdani}
\email{shokouhhaghdani@gmail.com} \affiliation{Department of
Physics, Faculty of Sciences, University of Isfahan, 81744 Isfahan,
Iran}
\author{Mohammad Ali Shahzamanian}
\affiliation{Department of Physics, Faculty of Sciences, University
of Isfahan, 81744 Isfahan, Iran}

\begin{abstract}
We theoretically investigate the spin diffusion coefficient tensor
in the $A$ phase of liquid $^{3}He$ in term of quasiparticle
life-time by using the Kubo formula approach at low temperatures. In
general, the coefficient is a fourth rank tensor for the anisotropic
states and can be defined as a function of both normal component of
spin-current and magnetization. The quasiparticle life-time is
obtained by using the Boltzmann equation. We find that components of
the spin diffusion coefficient are proportional to $T^{-2}$ at low
temperatures. The normal components of spin current, hence, are
strongly diffusive and one can ignore the contribution of these
components to the stability of half quantum vortices $(HQVs)$ in the
equal-spin-pairing of $^{3}He-A$ state. Hence to make a stable
$HQV$, it is enough for one to consider weak interaction plus the
effects of Landau Fermi liquid.
\end{abstract}

\maketitle

\section{Introduction}
After the discovery of new phases of $^{3}He$ , a large amount of
theoretical effort has been done on the spin collective
excitations\cite{Leggeett1,Vollhardt}. One view to the liquid phase
of  $^{3}He$ contains two normal and superfluid parts which at
$3\times10^{-3}$ Kelvin the superfluid part starts to be occurred
mostly in triplet pairing\cite{Leggeett1,Vollhardt}. The spin
triplet pairing in the superconductor compound $Sr_{2}RuO_{4}$
bellow $1.5$ Kelvin is observed experimentally\cite{Mackenzie} like
the $A$ phase of superfluid $^{3}He$ . The main similarity between
$^{3}He-A$ and $Sr_{2}RuO_{4}$ phases is that only Cooper's pairs
correspond to pair spin projections $S_{z}=1$ and $-1$ or
equal-spin-pairing $(ESP)$ in these phases. The normal state
properties of $Sr_{2}RuO_{4}$ are  obtained  in terms of a quasi two
dimensional Fermi liquid quantitatively \cite{Mackenzie}  whereas
$^{3}He-A$ is a three dimensional Fermi liquid. This discrepanting
manifest itself on the structure of the gap energies of them. The
triplet pairing contains particles with the same spin directions
that phenomenon leads to spin current. In two fluid model, this spin
current includes normal and condensate (Cooper's pairs) components.
The spin current of Cooper's pairs leads to interesting and
important phenomena such as $HQVs$ in the $ESP$ state. Unlike common
vortices, the half-quantum vortices contain half-integer
multiplications of the flux quantum $\Phi_{0}=hc/2e$ .
\\Vakaryuk and Leggett\cite{Vakaryuk} represented a different analysis in the
equal-spin-pairing superfluid state for a $HQV$. They used a $BCS$
like wave function with a spin-dependent boost and predicted that in
the stability of  $HQV$ an effective Zeeman field exists. In the
thermodynamic stability state, the effective Zeeman field produces a
non-zero spin polarization in addition to the polarization of
external magnetic field. In their theory the effect of diffusive
flow of the normal component did not take in to account. In this
paper we show that only at very low temperature this is the case and
in some limits of temperature the normal component might be not
diffusive and take part in stability of $HQV$.

\section{Theoretical approach}
Our formulation on spin diffusion coefficient is mainly concentrated
on the Kubo formula approach\cite{Shahzamanian1}. For the $A$ phase
of $^{3}He$ superfluid it is briefly written in the following.In
general, the spin diffusion $D$ is  a fourth rank tensor which may
be explained by the relation:
\begin{equation}\label{1}
\textbf{J}_{i\alpha}^{(n)}=\sum_{\beta j
l}D^{\alpha\beta}_{ij}(\frac{\partial}{\partial
x_{\beta}})(\textbf{S}_{j}-\chi_{jl}\textbf{H}_{l}),
\end{equation}
where $\textbf{J}_{n}, \textbf{S}, \textbf{H}$ and $\chi$ are the
spin current dyadic of the normal component, the magnetization, the
static magnetic field, and the static magnetic susceptibility
tensor, respectively. For the moment the dipole forces are ignored
and according to the conservation law one can write:
\begin{equation}\label{2}
(\frac{\partial\textbf{S}_{i}}{\partial
t})+(\frac{\partial}{\partial x_{\alpha}})\textbf{J}_{i\alpha}=0,
\end{equation}
in which the components of the total spin current $\textbf{J}$ are
\begin{eqnarray}\label{3}
\textbf{J}_{i\alpha}=\textbf{J}^{(n)}_{i\alpha}+\textbf{J}^{(s)}_{i\alpha}\;\;\;
\textbf{and}
\;\;\;\textbf{J}^{(s)}_{i\alpha}=(\frac{1}{2m})\rho^{\alpha\beta}_{ij}\Omega_{j\beta},
\end{eqnarray}
here $\Omega_{j\beta}$ is the component of the spin superfluid
velocity dyadic\cite{Shahzamanian1}. The dynamic susceptibility
tensor is defined by:
\begin{equation}\label{4}
\textbf{S}(\textbf{k},\omega)=\sum_{j}\chi_{ij}(\textbf{k},\omega)\textbf{H}_{j}(\textbf{k},\omega).
\end{equation}

By substituting Eq. (\ref{3}) and Eq. (\ref{1}) in Eq. (\ref{2}),
and to compare with Eq. (\ref{4}) is obtained:
\begin{equation}\label{5}
\lim_{\omega\rightarrow0}\lim_{k\rightarrow0}\frac{\omega}{k^{2}}\chi^{''}_{il}(\textbf{k},\omega)=\sum_{\alpha\beta
j}D^{\alpha\beta}_{ij}\widehat{k}_{\alpha}\widehat{k}_{\beta}\chi_{jl},
\end{equation}
here $\chi^{''}_{il}(\textbf{k},\omega)$ is the imaginary part of
the spin dynamic susceptibility  $\chi_{il}(\textbf{k},\omega)$.
Following Kadanoff and Martin's procedure on the imaginary part of
the dynamic susceptibility with the anti-commutator of the
magnetization in the normal phase one may generalize their result to
the anisotropic superfluid states\cite{Shahzamanian1}. In the absent
of the dipole interaction one may write:
\begin{equation}\label{6}
\langle\{S_{i}(\textbf{r},t),S_{j}(\textbf{r}^{'},t^{'})\}\rangle=\int\frac{d\omega}{\pi}\int\frac{d\textbf{k}}{(2\pi)^{3}}coth(\frac{\beta\omega}{2})\chi^{''}_{ij}(\textbf{k},\omega)\times\exp[i\textbf{k}.(\textbf{r}-\textbf{r}^{'})-i\omega(t-t^{'})].
\end{equation}\

With suppose that only the spin current associated with the normal
component takes part in the diffusive flow, the effects of the
superfluid component being negligible. By using Eq. (\ref{6}) and
Eq. (\ref{2}) in Eq. (\ref{5}), is obtained
\begin{equation}\label{7}
D^{\alpha\beta}_{ij}\chi_{jl}=\frac{1}{4}\beta\lim_{\omega\rightarrow0}\int^{\infty}_{0}dr\int
dt\ e^{i\omega t}\langle\{\textbf{J}_{i\alpha}^{(n)}(t),
\textbf{J}_{l\beta}^{(n)}(0)\}\rangle,
\end{equation}
$\langle \rangle$ represents the expectation value in the
equilibrium ensemble. By introducing the correlation function
$K(\tau)$ as $K^{\alpha\beta}_{ij}(\tau)=\langle
T_{\tau}\{\textbf{J}^{(n)}_{i\alpha}(\tau),\textbf{J}^{(n)}_{j\beta}(0)
\}\rangle$ in which the time $\tau$ varies between $-\beta$ to
$+\beta$. The spin diffusion coefficient $D$ may  be written in
terms of the correlation function\cite{Shahzamanian1};

\begin{equation}\label{8}
D^{\alpha\beta}_{ij}\chi_{jl}=\frac{1}{4}\lim_{\omega\rightarrow0}\frac{Im
K^{\alpha\beta}_{il}(\tau)(k=0,
i\omega_{l}\rightarrow\omega+i\eta)}{\omega},
\end{equation}
where
\begin{eqnarray}\label{9}
\nonumber K^{\alpha\beta}_{ij}(\omega_{n},
\textbf{k})=\frac{\mu^{2}}{8
m^{2}}\int\frac{d^{3}\textbf{p}}{(2\pi)^{3}}\frac{1}{\beta}\sum_{m}
e^{\xi_{m}\eta}(k_{\alpha}+2P_{\alpha})+(k_{\beta}+2P_{\beta})\times
\\Tr[\alpha^{i}G(\textbf{k}+\textbf{P},
\omega_{n}+\xi_{m})\alpha^{j}G(\textbf{P}, \xi_{m})].
\end{eqnarray}
in which $\alpha^{i}$ are the Pauli matrices in the four-dimensional
representation. After a bit of algebra Eq. (\ref{8}) can be written
as:

\begin{eqnarray}\label{10}
\nonumber
D^{\alpha\beta}_{ij}\chi_{jl}=\frac{\beta\chi_{n}V_{F}^{2}(1+\frac{1}{4}Z_{0})}{32}\int
d\xi\frac{d\Omega}{4\pi}\frac{d\omega}{2\pi}\widehat{p}_{\alpha}\widehat{p}_{\beta}\sec
h^{2}(\frac{\beta\omega}{2})\times \\Tr
\{\alpha^{i}[G(\textbf{P},\omega^{+})-G(\textbf{P},\omega^{-})]
\alpha^{l}[G(\textbf{P},\omega^{+})-G(\textbf{P},\omega^{-})]\}.
\end{eqnarray}

For the $^{3}He-A$ phase, the Green's function may be written as
\begin{equation}\label{11}
G(p,\omega)=\frac{\omega
Z_{p}(\omega)+\xi\rho_{3}\times1-Z_{p}(\omega)\Delta
(\mathbf{\alpha} .\textbf{d}) \sigma_{2}\times \rho_{2}
}{\omega^{2}Z^{2}_{p}(\omega)-\xi^{2}_{p}-Z^{2}_{p}(\omega)\Delta^{2}(\Omega)},
\end{equation}
where $Z_{p}(\omega)$ is the re-normalization function. We define
$\textbf{d}\equiv-\frac{1}{2}i\sum_{\alpha,\beta}(\sigma_{2}\sigma)_{\alpha\beta}\Delta_{\alpha\beta}$
and use $G(p,\omega)$ for obtaining $D^{\alpha\beta}_{ij}\chi_{jl}$;

\begin{equation}\label{12}
D^{\alpha\beta}_{ij}\chi_{jl}=\beta\chi_{n}V^{2}_{F}(1+\frac{Z_{0}}{4})\int\frac{d\Omega}{4
\pi}\widehat{p}_{\alpha}\widehat{p}_{\beta}\int^{\infty}_{\Delta}d\omega\frac{\sec
h^{2}(\frac{\beta\omega}{2})}{8Z_{2}(\omega^{2}-\Delta^{2})^{\frac{3}{2}}}\times[(2\omega^{2}-\Delta^{2})\delta_{il}+\Delta^{2}d_{i}d_{l}],
\end{equation}
here $Z_{2}$ is the imaginary part of $Z_{p}(\omega)$ and is simply
handled by using the poles of the single-particle Green's function
to define the quasiparticle energy and lifetime. Hence to lowest
order in the imaginary parts we may write
\begin{equation}
Z_{2}=\frac{EZ_{1}(E)}{2\tau(E,T)(E^{2}-\Delta^{2})}.
\end{equation}

Therefor, the spin diffusion coefficients for the $^{3}He-A$ phase
are written as
\begin{equation}\label{13}
D^{\alpha\beta}_{ij}\chi_{jl}=\chi_{n}V^{2}_{F}(1+\frac{Z_{0}}{4})\int\frac{d\Omega}{4
\pi}\widehat{p}_{\alpha}\widehat{p}_{\beta}\int^{\infty}_{0}d\xi\sec
h^{2}(\frac{\beta
E}{2})\times\frac{\beta\tau(E,T)}{4Z_{1}(E)}[2\delta_{il}-\frac{\Delta^{2}}{E^{2}}(\delta_{il}-d_{i}d_{l})].
\end{equation}

The above formula has been obtained with the condition
$\omega_{L}\tau_{D}<1$ where $\omega_{L}$ is the Larmor frequency
and $\tau_{D}$ is the spin diffusion lifetime. As to the
experimental values of $\omega_{L}$ \cite{Wheatley} this condition
is fulfill even at very low temperatures. Hence, we may compute
$\tau(E,T)$ at low temperatures. In this limit of temperature only
the quasiparticles which are located at the nodes of the gap energy
in the $^{3}He-A$ state take part to scattering processes and we may
write \cite{Shahzamanian2}
\begin{equation}\label{14}
\tau^{-1}(E,T)=(\frac{\pi\theta^{2}_{m}}{256\varepsilon_{F}})A^{2}_{s}\frac{[(\pi
k_{B}T)^{2}+E^{2}]}{[1+\exp(-\frac{E}{k_{B}T})]}
\end{equation}
where \cite{Greaves} $\theta_{m}\simeq\frac{\pi k_{B}T}{\Delta(0)}$
,$A_{s}=\sum_{l}S_{l}$ ,$S_{l}=A^{s}_{l}-3A^{a}_{l}$ and
$A^{s,a}_{l}=\frac{F^{s,a}_{l}}{[1+\frac{F^{s,a}_{l}}{(2l+1)}]}$,
where $F^{s,a}_{l}$ are the Landua's parameters\cite{Shahzamanian2}.

Finally, the spin diffusion coefficient in $^{3}He-A$ phase at low
temperatures are $D^{zz}_{zz}=D^{zz}_{xx}\simeq3D^{zz}_{yy}=C/T^{2}$
in which $C\simeq
\frac{16V^{2}_{F}(1+F_{0}^{a})\varepsilon_{F}\hbar}{A^{2}_{s}\pi^{2}k^{2}_{B}}$.

As it can be understood at low temperatures, the spin diffusion
coefficients increase as $T^{-2}$ and the normal components of spin
current are strongly diffusive and one can ignore the contribution
of these components to the stability of $HQV$.
\section{Conclusions}
In conclusion, we have investigated the temperature dependance of
spin diffusion coefficient of superfluid $^{3}He-A$ at low
temperatures. We have employed the Kubo approach and derived the
spin-diffusive coefficient of A phase of the Liquid $^3He$ in term
of quasiparticle life-time. The quasiparticle life-time is obtained
by using the Boltzmann equation. We have found that the
spin-diffusive coefficients at low temperatures regime are
proportional to $1/T^2$. The finding suggests strongly diffusive
normal components of spin-current and then weak contribution of them
to the stability of the half-quantum vortex state. Our work of
agreement with the assumption is in the recent work by Vakaryuk and
Leggett \cite{Vakaryuk}. The authors study the half-quantum vortex
phenomenon in the absence of normal components of spin-current at
$T=0$ and show the possibility of formation of the half-quantum
vortices in the equal-spin paring of Landau Fermi liquid. The normal
components of spin-current are important quantities which might
generate important influences and our findings show good agreement
with their assumption regarding the spin-current.


\end{document}